# A decision-support method for information inconsistency resolution in direct modeling of CAD models

Qiang Zou, Hsi-Yung Feng*
Department of Mechanical Engineering
The University of British Columbia
Vancouver, BC
Canada V6T 1Z4

**Abstract**

Direct modeling is a very recent CAD paradigm that can provide unprecedented modeling flexibility. It, however, lacks the parametric capability, which is indispensable to modern CAD systems. For direct modeling to have this capability, an additional associativity information layer in the form of geometric constraint systems needs to be incorporated into direct modeling. This is no trivial matter due to the possible inconsistencies between the associativity information and geometry information in a model after direct edits. The major issue of resolving such inconsistencies is that there often exist many resolution options. The challenge lies in avoiding invalid resolution options and prioritizing valid ones. This paper presents an effective method to support the user in making decisions among the resolution options. In particular, the method can provide automatic information inconsistency reasoning, avoid invalid resolution options completely, and guide the choice among valid resolution options. Case studies and comparisons have been conducted to demonstrate the effectiveness of the method.

## 1. Introduction

Computer-aided design (CAD) is a widely used tool in mechanical design practices, including automotive, shipbuilding, and aerospace industries. The most recent CAD paradigm is called direct modeling, whose main feature is to allow users to interact directly with the geometry of the model to make edits [1,2]. Typically, to modify a model, users just need to grab, push, and pull the geometric entities of interest in the model, see [3] for some examples. Despite the excellent modeling flexibility, direct modeling lacks the parametric editing capability as it focuses on pure solid models without any built-in associativity information. The ability to attain model variants via parametric edits has been identified by many authors as an essential feature of modern CAD systems [4–7]. New developments are thus necessary for direct modeling to have the parametric capability.

To achieve the goal, associativity information needs to be incorporated into direct modeling. One may alternatively incorporate direct modeling into parametric (feature) modeling to attain both direct and parametric capabilities, as done by some CAD vendors [2] and in [8], but this way has inherent limitations such as leading to a restricted direct modeling functionality [9,10]. Incorporating associativity information into direct modeling is essentially to wrap a geometric constraint system (GCS) around the solid model direct modeling focuses on. Parametric edits can then be made through the associativity (GCS) information layer, and direct edits can be made via the geometry (solid model) information layer. The major issue here is: when an information layer is edited, the changes are not reflected in the other automatically. As a result, the consistency of the two information layers in the pre-edit model is broken, and an invalid model is generated.

There are two types of geometry-associativity inconsistency (GAI). When parametric edits are made to the model, the GCS is changed and becomes inconsistent with the unchanged geometry, which states the first type. Resolving inconsistencies of this kind mainly involves a re-evaluation of the changed GCS. This can be well-addressed by existing work in the area of geometric constraint solving [10]. The second inconsistency type refers to situations where the geometry changed by direct edits becomes inconsistent with the unchanged GCS. Such inconsistencies take the form of under-constrained and over-constrained parts in the model, necessitating the maintenance of a well-constrained model (Section 3 will give detailed descriptions on this). The major issue of resolving these inconsistencies is that there often exist many resolution options. Some of them will lead to invalid (non-well-constrained) models; among the valid models, there will only be one in line with the designer's intent. As a result, the resolution process is prone to invalid and unintended modeling results, and thus a challenging problem. In the following, GAI refers exclusively to the second type, unless otherwise stated.

* Corresponding author. Tel.: +1-604-822-1366; fax: +1-604-822-2403.
E-mail: john.qiangzou@gmail.com (Q. Zou), feng@mech.ubc.ca (H.-Y. Feng)

Design intent is generally too complicated to infer satisfactorily by the computer [11,12]. As a result, a completely automatic GAI resolution method may not be possible at least at present, and a decision-support scheme may be a more practical choice. A good decision-support scheme should leave the least degrees of intellectual effort to the user. GAI resolution consists primarily of two tasks: reasoning inconsistencies and making decisions among resolution options. The former task is to know what inconsistencies a model to be resolved has and to understand how they are formed. The challenge of this task lies in taking out and decoupling the inconsistencies, which can be translated to the following problem: decompose the model into minimal over-constrained parts and maximal well-constrained parts[1]. An automatic inconsistency reasoning method that can effectively address these problems is to be presented in this work. The main benefit of this method is that it allows a complete exclusion of invalid resolution options.

Then the latter decision-making task becomes much easier as the remaining resolution options are all valid. In other words, the user does not need to worry about generating invalid modeling results. To further facilitate the choice among valid resolution options, this work prioritizes them and then presents the prioritized options incrementally to the user for acceptance or rejection. This will much reduce the required intellectual effort for GAI resolution. The main feature of the prioritization is that it allows a resolution option leading to a smaller model variation to occur first in the final prioritization. Eventually, a decision-support method can be developed for GAI resolution, with guaranteed model validity and minimal model variation (up to the user's discretion).

## 2. Related work

Publications related to this work can be classified into three topics and are primarily from two research areas. The topics are (1) constraint state (under-, well-, or over-constrained) characterization (2) optimal model decomposition and (3) constraint prioritization. The two research areas are geometric constraint solving and CAD model beautification.

For the first topic, many methods have been presented in the last three decades, mainly in the area of geometric constraint solving. A thorough review of them can be found in [10]. Among the many methods, the category of interest to this work is called the witness configuration method as it has the following advantages: (1) it is mathematically sound [13]; (2) it is general and has no limitation on model representation schemes [9]; and (3) it has successful real applications/validations [14,15]. The witness configuration method was first proposed in [16] and later detailed in [9,13,17,18]. It is based on the property that models of different constraint states have different behavior under infinitesimal perturbations made to the model geometry. Formalizing the different behavior yields the mathematical criteria for constraint states. In this work, this method is to be used for constraint state characterization.

For the second topic, most publications are also from the area of geometric constraint solving. The methods related to the witness configuration method include [13–15,17]. Although presented in different forms, these methods share the same idea: greedy algorithms have been used to attain the minimal over-constrained parts and maximal well-constrained parts in a model. As known, greedy algorithms could, however, fail to generate optimal solutions (and therefore undesirable decomposition results). In fact, such failure cases have been observed, as will be shown in Section 5. It is safe to say that the use of greedy algorithms does not give a formulation of the decomposition problem but represents an incomplete technical tool. This work will present a new, precise formulation of the decomposition problem.

For the third topic, related work comes primarily from both of the two previously mentioned areas. In the geometric constraint solving area, the early attempt may be made by [19–21]. They are graph-based and cannot handle models having constraint dependencies (except for the simplest structural dependencies) [16], making them inapplicable to this work. Due to this limitation, recent studies have shifted to the witness configuration method [14,15]. The developed methods are, however, restricted to dealing merely with over-constrained models, which makes them inapplicable to the general models — having both under-constrained and over-constrained parts — considered in this work.

The CAD model beautification area collects studies of removing imprecisions (and therefore called beautification) in reverse engineered models through the guidance of geometric constraints. Constraint prioritization is needed because we need to decide which constraints to use for the beautification. There are two classes of methods in this area: qualitative and quantitative. The qualitative scheme prioritizes constraints based on their types and a set of heuristics [22–25]. Such methods cannot handle prioritization among constraints having the same type. This is where the quantitative scheme can help. Existing quantitative methods [26,27] prioritize constraints according to the deviation of the nominal parameter value of a constraint from its corresponding dimension in the reverse engineered model. Apparently, this kind of prioritization only works if there are imprecisions in the model, which is not the case for this work. Nevertheless, the qualitative-then-quantitative strategy has been seen to be useful for constraint prioritization and is to be followed by this work, with a new quantitative prioritization scheme.

[1] Attaining maximal well-constrained parts is equivalent to getting minimal under-constrained parts since under-constraint is described by degrees of freedom between well-constrained parts.

The above review suggests that the documented studies on the problem of GAI resolution are insufficient: only small part of the problem can be solved using existing work, but the other cannot. New/improved methods are to be presented in this paper to address this insufficiency and to attain effective decision-support for GAI resolution.

## 3. Issues of geometry-associativity inconsistency resolution

This section explains the formation of GAI and the challenge of GAI resolution. When a direct edit is made to a model, the geometry information (i.e., a boundary representation solid model [28]) is to be updated. The resulting solid model, or a portion thereof, will have new boundary faces and dimensions, leading to disagreements with the constraints in the pre-edit model GCS, see the second and third columns in Fig. 1 for examples. Hence, the model GCS needs to be updated accordingly: replace the parameter values of the constraints with the dimensions of the new solid model, and remove any inapplicable constraints, as exemplified by the fourth column in Fig. 1.

| Pre-Edit Model | | Updated Model | Updated Model | Constraint |
|---|---|---|---|---|
| Geometry | GCS | Geometry | GCS | State Change |
| 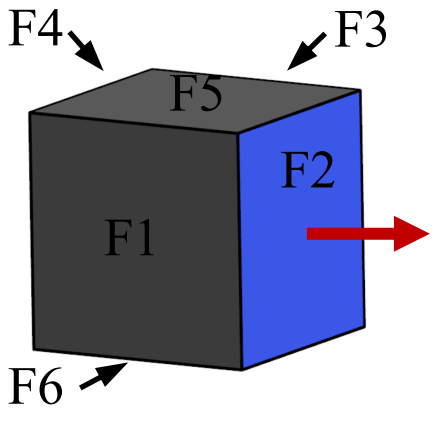  | 1. Distance(F1,F3)=1<br>2. Distance(F2,F4)=1<br>3. Distance(F5,F6)=1<br>4. Perpendicular(F1,F5)<br>5. Perpendicular(F1,F4)<br>6. Perpendicular(F4,F5) | 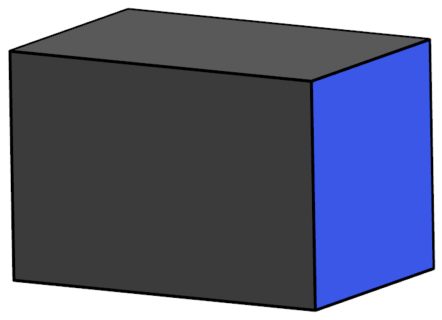 | 1. Distance(F1,F3)=1<br>2. **Distance(F2,F4)=2**<br>3. Distance(F5,F6)=1<br>4. Perpendicular(F1,F5)<br>5. Perpendicular(F1,F4)<br>6. Perpendicular(F4,F5) | Well To Well |
| 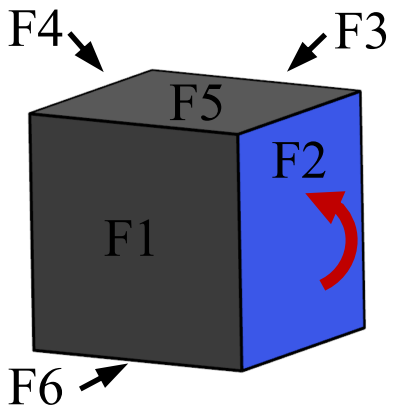  | 1. Distance(F1,F3)=1<br>2. Distance(F2,F4)=1<br>3. Distance(F5,F6)=1<br>4. Perpendicular(F1,F6)<br>5. Perpendicular(F1,F4)<br>6. Perpendicular(F4,F6) | 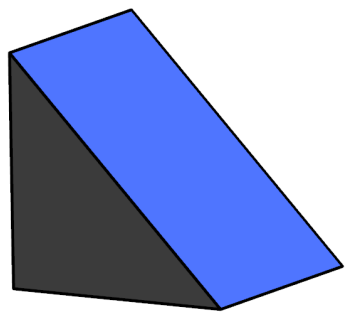 | 1. Distance(F1,F3)=1<br>2. **(Removed)**<br>3. **(Removed)**<br>4. Perpendicular(F1,F6)<br>5. Perpendicular(F1,F4)<br>6. Perpendicular(F4,F6) | Well To Under |
| 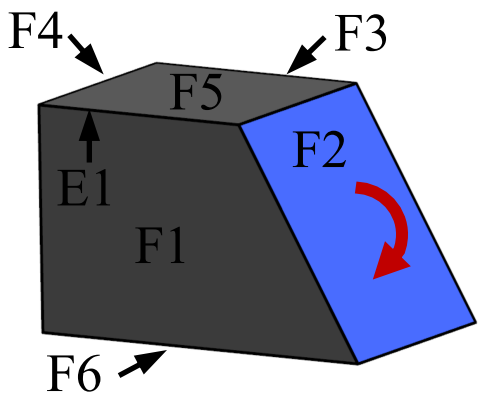  | 1. Distance(F1,F3)=1<br>2. Distance(F5,F6)=1<br>3. Perpendicular(F1,F5)<br>4. Perpendicular(F1,F4)<br>5. Perpendicular(F4,F5)<br>6. Angle(F2,F4)=150°<br>7. Angle(F2,F5)=60°<br>8. Length(E1)=1 | 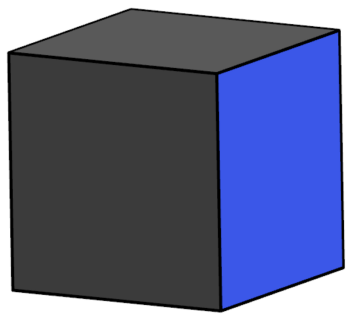 | 1. Distance(F1,F3)=1<br>2. Distance(F5,F6)=1<br>3. Perpendicular(F1,F5)<br>4. Perpendicular(F1,F4)<br>5. Perpendicular(F4,F5)<br>6. **Parallel(F2,F4)**<br>7. **Perpendicular(F2,F5)**<br>8. Length(E1)=1 | Well To Over |

Figure 1: Direct edits, GCS updates, and varying constraint state change results (blue faces: push-pulled faces; straight arrow: translational push-pull; curved arrow: rotational push-pull).

The GCS update process above is simple in its own right, but it can lead to varying results. If a valid model (being well-constrained) is output as in the top example of Fig. 1, nothing further needs to be done. If otherwise, there are inconsistencies between the geometry information and associativity information in the model, and GAI resolution becomes a necessity. These inconsistencies take the form of under-constraint and over-constraint. Under-constraint means that there are fewer constraints in the model GCS than needed to fully restrict the model geometry, and over-constraint means that there are more constraints than needed. The middle example in Fig. 1 shows an under-constraint situation (plane F2 has free motions with respect to other planes). The bottom example in Fig. 1 shows an over-constraint situation (constraints 5, 6 and 7 are dependent).

In GAI resolution, the applied method should always output valid models. This is, however, not straightforward because there exist many resolution options. Consider the bottom example in Fig. 1. The over-constraint involves a cyclical dependency among constraints 5, 6 and 7. Removing any constraint not involved in the dependency (i.e., constraints 1, 2, 3, 4 or 8) cannot resolve the over-constraint and leads to a failed resolution. Only removing a constraint relevant to the dependency can resolve the over-constraint successfully.

From the same example, we can also see that the generation of valid modeling results lies in the detection of over-constrained parts and under-constrained parts. That is, if the over-constrained part (i.e., constraints 5, 6 and 7) is known, invalid resolution options (i.e., constraints 1, 2, 3, 4 or 8) can be excluded, and then chances of generating invalid modeling results can be eliminated. The same applies to under-constraint situations: given degrees of freedom (representing under-constraint) of the model, invalid resolution options can be effectively excluded. Over-constrained and under-constrained parts should be attained in a way that a part is irreducible to smaller parts in order to decouple the parts and to make reasoning individual inconsistencies easy. A part being irreducible to smaller parts can be stated mathematically as: the part is of minimal size. For under-constrained parts, minimizing them is equivalent to maximizing well-constrained parts, as already noted in footnote #1.

Even the applied resolution method successfully avoids invalid resolution options, there are often more than one valid resolution options. The options in Table 1 are typical examples of such a situation. To support the choice among valid resolution options, one viable way is to prioritize them and then to recommend them to the user incrementally. An effective prioritization scheme should give a good measure of the impact of applying a resolution option — removing/adding a constraint from the model GCS — on the model geometry. This problem is, however, not trivial as the qualitative operation of removing/adding constraints has no direct connections to the quantitative notion of model geometry.

In summary, GAI takes the form of over-constrained and under-constrained parts. The challenge of GAI resolution lies in effectiveness towards detecting minimal over-constrained parts and maximal well-constrained parts in the model, as well as an effective criterion for prioritizing valid resolution options.

Table 1: Valid resolution options for the bottom example in Fig. 1.

| Updated Model GCS | Resolution Option 1 | Resolution Option 2 | Resolution Option 3 |
|---|---|---|---|
| 1. Distance(F1,F3)=1 | 1. Distance(F1,F3)=1 | 1. Distance(F1,F3)=1 | 1. Distance(F1,F3)=1 |
| 2. Distance(F5,F6)=1 | 2. Distance(F5,F6)=1 | 2. Distance(F5,F6)=1 | 2. Distance(F5,F6)=1 |
| 3. Perpendicular(F1,F5) | 3. Perpendicular(F1,F5) | 3. Perpendicular(F1,F5) | 3. Perpendicular(F1,F5) |
| 4. Perpendicular(F1,F4) | 4. Perpendicular(F1,F4) | 4. Perpendicular(F1,F4) | 4. Perpendicular(F1,F4) |
| 5. Perpendicular(F4,F5) | 5. Perpendicular(F4,F5) | 5. **(Removed)** | 5. Perpendicular(F4,F5) |
| 6. Parallel(F2,F4) | 6. **(Removed)** | 6. Parallel(F2,F4) | 6. Parallel(F2,F4) |
| 7. Perpendicular(F2,F5) | 7. Perpendicular(F2,F5) | 7. Perpendicular(F2,F5) | 7. **(Removed)** |
| 8. Length(E1)=1 | 8. Length(E1)=1 | 8. Length(E1)=1 | 8. Length(E1)=1 |

## 4. Methodology

### 4.1. The framework

The GAI resolution framework used in this work is shown in Fig. 2. It begins with a well-constrained model and a direct edit applied to it, then performs GCS update according to the new model geometry, then sends the updated model to an analyzer (the modules in diamond shape) to evaluate its constraint state. If the model is still well-constrained, nothing further needs to be done; if not, the analyzer directs the workflow to different inconsistency resolution branches. In both branches, it first takes out information inconsistencies (the two detection modules) and then, based on the detection results, generates and prioritizes valid resolution options (the two prioritization modules), then presents the prioritized options to the user for decisions. The model after resolution is sent again to the analyzer for making sure that the model has become well-constrained. This last procedure is necessary because the user can do anything beyond what is suggested.

Essential modules in this framework are the analyzer, detection, and prioritization modules. Their implementation methods are to be presented in the following subsections, respectively. In particular, the detection and prioritization modules address directly the GAI resolution challenge stated in the previous section.

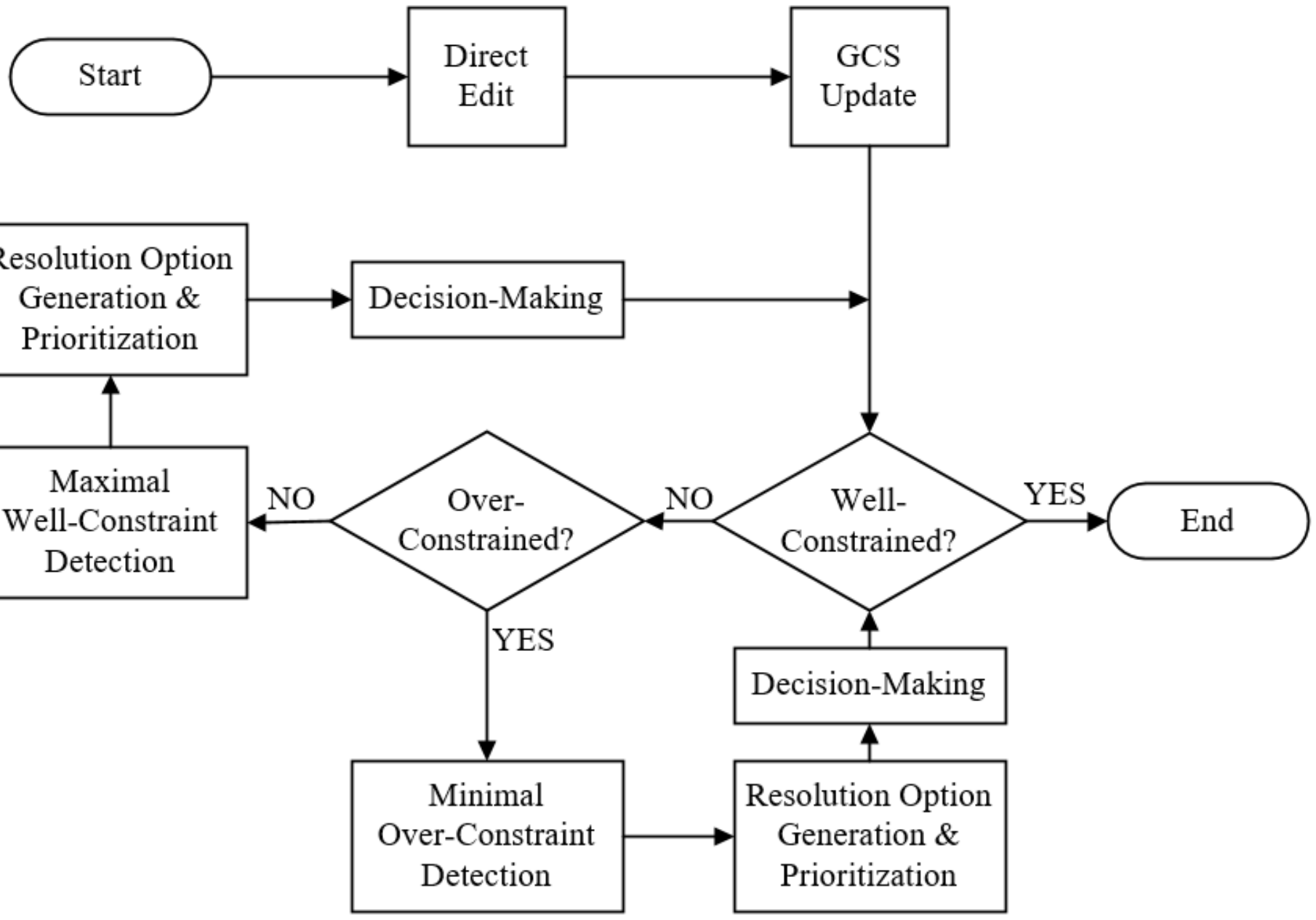

Figure 2: GAI resolution framework.

### 4.2. The analyzer module

The task of this module is to check if there are over-constrained and/or under-constrained parts in the input model. The output is just a yes or no; no further information about the parts are needed at this stage. The authors have previously presented a witness configuration method to characterize a model's constraint state [9]. The method is applicable to general models and has a proof of correctness. It is summarized below.

The input model's GCS can be described mathematically by a system of algebraic equations, denoted by $F(X) = 0$ where the variable $X$ represents coordinates of geometric entities in the model geometry. The witness configuration method examines how the equations $F(X)$ behave under infinitesimal perturbations made to the variable $X$. Let $\Delta X$ represent a perturbation. The corresponding change made to the equations is given by:

$$\Delta F = J(X)\Delta X + O(\|\Delta X\|_2^2) \tag{1}$$

where $J(X)$ is the Jacobian matrix evaluated at point $X$. Here, the perturbation $\Delta X$ is parametric and does not carry any geometric meanings explicitly. This issue can be solved by using a transformation matrix $T$ to relate the parametric perturbation $\Delta X$ with its geometric counterpart $\Delta X'$ that expresses perturbations in terms of geometric motions (translations and/or rotations) as follows: $\Delta X = T\Delta X'$. (The construction of matrix $T$ can be found in Section 3.2, [9].) Then the relationship between a geometric perturbation $\Delta X'$ and the equation change $\Delta F$ is:

$$\Delta F = J(X)T\Delta X' + O(\|T \cdot \Delta X'\|_2^2) \tag{2}$$

For simplicity, we use a single matrix $G$ to represent the product $J(X)T$ and call it the geometric perturbation matrix.

Models of different constraint states have different change patterns for $\Delta F$ under same perturbations, which ultimately leads to different structures on the geometric perturbation matrix. To be more specific, a model containing over-constrained parts yields a matrix $G$ having linearly dependent rows. A model having under-constrained parts gives a matrix $G$ whose null space is bigger than the nominal free perturbation space. The nominal free perturbation space consists of perturbations that do not change the model geometry, e.g., rigid-body motions. Following immediately, the geometric perturbation matrix of a well-constrained model has no dependent rows, and its null space equals the nominal free perturbation space.

It should be noted that the statements above are valid only when the geometric perturbation matrix $G$ is evaluated at a carefully selected point called the witness configuration where certain types of constraints has already been satisfied by the geometry undergoing perturbation [13]. This requirement is trivial for this work: due to the prior procedure of GCS update, all the constraints in the input model's GCS agree with the input model's geometry, and this geometry thus serves as a perfect witness configuration.

The formal checking conditions for different constraint states are as follows. A model is over-constrained if

$$NullSpace(G^T) \neq \emptyset \tag{3}$$

A model is under-constrained if

$$NullSpace(G) \supset N \tag{4}$$

where $N$ denotes the nominal free perturbation space (see Section 3.3, [9] for its construction). A model is well-constrained if

$$\begin{aligned} &NullSpace(G^T) = \emptyset \\ &NullSpace(G) = N \end{aligned} \tag{5}$$

### 4.3. The detection module

The tasks of this module are to attain minimal over-constrained parts and/or maximal well-constrained parts in the input model. Its outputs are to be used as inputs for generating (valid) resolution options. Besides, these tasks are the building blocks of achieving an automatic inconsistency reasoning method. Different from the constraint state characterization problem in Section 4.2, the detection problem here has not yet been clearly formulated, and there is a lack of effective methods. This issue is to be addressed in this subsection.

#### 4.3.1. Minimal over-constraint detection

First of all, the term minimal over-constraint should be made formal. An over-constrained part of the input model takes the form of a group of constraints having dependencies. An over-constrained part is minimal if its size is minimized. Such a part only consists of constraints relevant to the dependency; no irrelevant constraints will be included.

As already noted, constraint dependencies yield linearly dependent rows in the geometric perturbation matrix. To be precise, a vector $x \in NullSpace(G^T)$ represents a dependency group, and the nonzero elements of $x$ indicate the constraints involved in this group. A dependency group being minimal is thus to say that the vector $x$ has the minimal number of nonzero elements, which can be modeled as:

$$\min_{x} \|x\|_0 \quad s.t. \quad G^T x = 0,\ x \neq 0 \tag{6}$$

where $\|\cdot\|_0$ is the $\ell_0$ norm whose mathematical meaning is to count nonzero elements in a vector. A minimal dependency group should be irreducible to smaller dependency groups, meaning all dependency vectors $\{x_i\}_{i=1}^{n}$ should be linearly independent. Suppose we have got the first $k < n$ dependency vectors $\{x_i\}_{i=1}^{k}$. The requirement that the next dependency vector is linearly independent with $\{x_i\}_{i=1}^{k}$ can be modeled as:

$$\min_{x_{k+1}} \|x_{k+1}\|_0 \quad s.t. \quad G^T x_{k+1} = 0,\ x_{k+1} \neq Span(x_1 \cdots x_k) \tag{7}$$

The modeling in (7) suggests that dependency vectors are to be attained sequentially. Solving the optimization problem in (7) is not easy due to its non-convexity. In the following, this optimization problem will be reformulated to a typical sparse recovery (a.k.a. compressive sensing) problem, through a series of mathematical manipulations. With the reformulation, the problem will become a tractable optimization problem using existing optimization techniques. The reformulation is only of mathematical interest and does not change the essence of the modeling expressed in (7) or provide new insights into the problem.

Divide the basis of $NullSpace(G^T)$ into two parts:

$$NullSpace(G^T) = \Big[\overbrace{x_1 \ldots x_k}^{A} \quad \overbrace{A's\ orthogonal\ complement}^{B}\Big] \tag{8}$$

Then any vector $x \in NullSpace(G^T)$ can be represented as a linear combination of the column vectors of $A$ and $B$, denoted by $Ay + Bz$ where $y, z$ are coefficient vectors. The constraints of the optimization problem in (7) then takes the form of $x_{k+1} = Ay + Bz$ and $z \neq 0$. The inequality $z \neq 0$ can be further translated to a condition on the vector $x_{k+1}$, as follows. The equation $x_{k+1} = Ay + Bz$ can be expressed in terms of the solution to the following optimization problem:

$$\min_{y,z} \|x_{k+1} - Ay - Bz\|_2^2 \tag{9}$$

The solution to this optimization problem is (Chapter 4, [29]):

$$\binom{y}{z} = \left((A \quad B)^T (A \quad B)\right)^{-1} (A \quad B)^T x_{k+1} \tag{10}$$

$z$ can then be expressed as:

$$z = (0 \quad I)\binom{y}{z} = (0 \quad I)\left((A \quad B)^T (A \quad B)\right)^{-1} (A \quad B)^T x_{k+1} \tag{11}$$

where $I$ is an identity matrix whose dimension is the same as the size of $B$'s columns. Simplify the notations in Eq. (11) to $z = Cx_{k+1}$, and constrain $z$ to lie on a unit sphere in order to avoid a vanished $z$. The optimization problem in (7) is then transformed into:

$$\min_{x_{k+1}} \|x_{k+1}\|_0 \quad s.t. \quad G^T x_{k+1} = 0, \ \| Cx_{k+1} \|^2 = 1 \tag{12}$$

which is a typical sparse recovery problem that can be effectively solved with the relaxation method [30].

With the formulation (12) in place, a sequential method of attaining minimal over-constrained parts can be made available, which is summarized in Algorithm 1. The algorithm attains individual parts by first solving the optimization problem in (12) then getting nonzero elements in the solved $x_i$ and finally mapping these elements to their corresponding constraints (Lines 4-6).

**Algorithm 1**: Minimal Over-Constraint Detection

**Input:** $G -$ the geometric perturbation matrix
**Output:** $P -$ minimal over-constrained parts

1. $P \leftarrow \emptyset$
2. $N \leftarrow \mathrm{Dim}\big(NullSpace(G^T)\big)$
3. **for** $i \leftarrow 1$ to $N$ **do**
4. $\quad x_i \leftarrow$ Solve the optimization problem in (12)
5. $\quad idx \leftarrow$ IndexOfNonZeroElements($x_i$)
6. $\quad C \leftarrow$ MapToConstriants($idx$)
7. $\quad P \leftarrow P \cup \{C\}$
8. **end for**
9. **Return** $P$

### 4.3.2. Maximal well-constraint detection

A well-constrained part refers to a subset of the model geometry whose induced subsystem from the model GCS is well-constrained. A part's induced subsystem is the subset of the constraints in the model GCS that are defined within the part. A well-constrained part $P$ is maximal if there is no another well-constrained part $P'$ such that $P \subset P'$.

From the framework presented in Fig. 1, it can be seen that the input model for the maximal well-constrained detection module contains no over-constrained parts. The well-constraint checking conditions in Eq. (5) can thus be reduced to only one condition: $NullSpace(G) = N$. A vector $x \in NullSpace(G)$ describes a free perturbation that does not violate any constraints in the model GCS due to the equality $\Delta F = Gx = 0$ (see Eq. (2)). Hence, the condition $NullSpace(G) = N$ essentially means that a well-constrained model's free perturbations are all nominal free perturbations. When it comes to a part of the model, the checking condition is approximately the same: any free perturbations of the geometric entities involved in the part are nominal free perturbations.

Next, we derive the mathematical condition for a part to be deemed a well-constrained part. Let $I$ denote the index set of the geometric entities in a part of interest, $f \in NullSpace(G)$ a free perturbation, and $\delta_i(f)$ a function getting the component in $f$ that corresponds to the $i$-th geometric entity. The part represented by $I$ is well-constrained if the perturbations $\{\delta_{I_1}(f) \cdots \delta_{I_m}(f)\}, \forall f \in NullSpace(G)$ are all nominal free perturbations. Whether a perturbation is a nominal free perturbation or not can be examined by seeing if it can be expressed as a linear combination of the basis vectors of the space $N$. Let $B$ represent a matrix whose columns are these basis vectors. The part is well-constrained if the following equations are solvable for any $f \in NullSpace(G)$:

$$\delta_i(Bx) - \delta_i(f) = 0, \quad i \in I \tag{13}$$

where $x$ is the variable and represents the linear combination coefficients.

If the part is maximal, any geometric entity $k \notin I$ should have the following property: there exists a free perturbation $f \in NullSpace(G)$ such that the perturbations $\{\delta_{I_1}(f) \cdots \delta_{I_m}(f), \delta_k(f)\}$ are not nominal free perturbations. To state mathematically, the following equations should have no solution for a certain $f \in NullSpace(G)$:

$$\begin{aligned} &\delta_i(Bx) - \delta_i(f) = 0, \quad i \in I \\ &\delta_k(Bx) - \delta_k(f) = 0 \end{aligned} \tag{14}$$

Or equivalently, any solution to the part $\delta_i(Bx) - \delta_i(f) = 0,\ i \in I$ leads to the inequality $\delta_k(Bx) - \delta_k(f) \neq 0$. This property is the key to attaining maximal well-constrained parts.

Suppose we want to find the largest well-constrained part in the input model. This part having the maximal size means that, in the following linear system, the number of equations that can be satisfied is maximal:

$$Bx - f = 0, \quad \forall f \in NullSpace(G) \tag{15}$$

Eq. (15) can be rewritten in the following matrix form:

$$BX - F = 0 \tag{16}$$

where $F$ is a matrix whose columns are the basis vectors of $NullSpace(G)$. Maximizing the number of satisfied equations in (16) is equivalent to minimizing the number of unsatisfied equations, that is:

$$\min_X \ number\ of\ nonzero\ rows\ of\ BX - F \tag{17}$$

Let $r_i$ be the $i$-th row of $BX - F$. $r_i$ is a nonzero row iff $\|r_i\|_2 = \sqrt{r_i \cdot r_i^T} \neq 0$ where $\|\cdot\|_2$ denotes a vector's $\ell_2$ norm. If we collect all the rows' $\ell_2$ norms into a vector, denoted by $r$, then the number of nonzero rows of $BX - F$ is the $\ell_0$ norm of $r$. This sequential application of two norms is often called the mixed $\ell_0/\ell_2$ norm, denoted by $\|BX - F\|_{2,0}$. The problem in (17) can then be modeled as:

$$\min_X \ \|BX - F\|_{2,0} \tag{18}$$

The essential part of this optimization problem is the $\ell_0$ norm. Then (18) is, again, a sparse recovery problem. Let $X^*$ be the minimizer of this optimization problem; the zero rows of the matrix $BX^* - F$ correspond to the largest well-constrained part in the model.

The second largest well-constrained part (and so forth) can be attained similarly. After the largest well-constrained part is attained, we can focus on the remaining geometric entities in the model, then update the matrices $B$ and $F$ accordingly, and finally solve the optimization problem in (18) again. The update of $B$ and $F$ can be done by removing the rows that correspond to the geometric entities in the known largest well-constrained part. Apparently, by repeating the above updating and solving procedures, all the maximal well-constrained parts can be attained sequentially, from the largest to the smallest. Algorithm 2 shows the procedures of doing so. In particular, Lines 7 and 8 exclude the detected well-constrained part from the model and store it as a new maximal well-constrained part.

**Algorithm 2**: Maximal Well-Constraint Detection

**Input:** $B, F$ − the nominal free perturbation basis and free perturbation basis
**Output:** $P$ − maximal well-constrained parts

1. $P \leftarrow \emptyset$
2. $M \leftarrow$ GetAllGeometricEntities()
3. **while** $|M| \neq 0$ **do**
4. Update $B$ and $F$
5. $X^* \leftarrow$ Solve the optimization problem (18)
6. $I \leftarrow$ IndexOfZeroRows($BX^* - F$)
7. $M' \leftarrow \{M(i)|\ i \in I\}$, $M \leftarrow M - M'$
8. $P \leftarrow P \cup \{M'\}$
9. **end while**
10. **Return** $P$

### 4.4. The prioritization module

The tasks of this module are to generate valid resolution options and prioritize these options in order to present them to the user incrementally. Similar to the situation for the detection module, there is a lack of effective methods for the generation and prioritization tasks here. This subsection serves to address this issue.

#### 4.4.1. Resolution Option Generation

The previous detection module has prepared the input model's over-constraint information in the form of groups of minimal dependent constraints. As the groups are minimized and cannot be decomposed into smaller subgroups, there is only one cyclical constraint dependency in each group. As a result, the removal of any constraint in a group can break the cyclical dependency in this group and resolve the associated over-constraint. In other words, the valid resolution options for resolving the over-constraint in a group are the group's constraints themselves.

When considering all the minimal over-constrained parts in the input model, resolution option generation should be done dynamically for them. One can imagine a situation in which two minimal over-constrained parts have a common constraint; if this constraint is chosen for resolving one part, the over-constraint in the other part is resolved automatically. Therefore, resolution option generation for all the parts should not be done parallelly. Instead, the generation-then-resolution process should be done one part by one part.

The given under-constraint information of the input model takes the form of maximal well-constrained parts. The degrees of freedom (DOFs) between these parts describe the model's under-constraint. As the parts are maximal, viable constraints to eliminate the DOFs are those bridging any two of the parts while not adding new over-constraint to the model. Any constraints that are defined within individual parts are invalid resolution options.

To attain valid resolution options, this work employs a two-step generation scheme. The first step is to generate a naïve constraint set consisting of all possible bridging constraints between two maximal well-constrained parts of interest. This is to be done by looking up a constraint table using the geometric entity pair from respective parts as the key. Any constraints that can be expressed by the constraint types in Table 2 will be generated and stored in the naïve constraint set. The second step is to remove constraints in the naïve constraint set that will cause over-constraint with existing constraints. The removal process is trivial due to the over-constraint checking condition made available in Section 4.2. We just need to use this condition to check every generated constraint.

The resolution options given by the two-step generation scheme above need to be updated dynamically. When one of the generated resolution options is chosen by the user and added to the model, some of the previously valid resolution options could become invalid due to the newly added constraint. To remove such resolution options, we just need to perform the second step described above again.

From the descriptions above, the generation of valid resolution options may seem easy. This is because the generation process is under a very advantageous situation where all information inconsistencies in the input model have been clearly isolated and decoupled. With this advantage, it becomes trivial to reason the formation of individual information inconsistencies. For example, every constraint in a minimal over-constrained part contributes to the associated constraint dependency; thereby, the formation of the information inconsistency is obvious. In this regard, the detection methods presented previously gives an automatic inconsistency reasoning method, which in turn allows an easy generation of valid resolution options.

Table 2: Geometric constraint look-up table.

| Geometric Constraint | Equation Representation |
|---|---|
| Angle $\alpha$ between directions $d_1, d_2$ | $d_1^T d_2 = cos\ (\alpha)$ |
| Parallel directions $d_1, d_2$ | $d_1 \times d_2 = 0$ |
| Distance $l$ between two positions $p_1, p_2$ | $\|p_1 - p_2\|_2 = l$ |
| Equal position $p_1, p_2$ | $p_1 = p_2$ |
| Equal angle parameters $\alpha_1, \alpha_2$ | $\alpha_1 = \alpha_2$ |
| Equal length parameters $l_1, l_2$ | $l_1 = l_2$ |

### 4.4.2. Resolution Option Prioritization

Resolution options take the form of geometric constraints that are to be added to or removed from the input model. Prioritizing them is to attain a permutation of them for putting them in a certain order. The final permutation is determined by a binary comparison operation that accepts two constraints as arguments and determines which of them should occur first in the final permutation. In this work, a two-level comparison scheme is employed, a hybrid of the existing work and a new method.

The first level is called the rough comparison and based on constraint types. Existing studies [22–25] have shown that certain types of constraints could carry more engineering knowledge and occur more frequently than others, and that a type-based comparison can be effective. For this reason, the type-based comparison method is also empolyed in the present work, and Table 3 shows the type precedence[2] used, which is an adaption from [25]. This kind of comparison has a limitation that it cannot handle situations where two constraints have the same type/precedence. This is the role to be played by the comparison at the second level, referred to as the fine comparison.

Table 3: Rough prioritization based on constraint type.

| **Geometric Constraint** | | | **Precedence (1: high, 5: low)** |
|---|---|---|---|
| **Entities** | | **Type** | |
| Face | Face | Parallel/perpendicular directions,<br>Distance between positions,<br>Equal size parameters | 1 |
| | | General angle between directions | 2 |
| Face | Edge | Parallel/perpendicular directions,<br>Distance between positions | 2 |
| | | General angle between directions | 4 |
| Face | Vertex | Distance between positions | 5 |
| Edge | Edge | Equal angle/length parameters,<br>Parallel/perpendicular directions,<br>Distance between positions | 3 |
| | | General angle between directions | 5 |
| Edge | Vertex | Distance between positions | 5 |
| Vertex | Vertex | Distance between positions | 5 |

The proposed fine comparison scheme is based on the following observation: parameter changes made to different constraints in a model often yield different degrees of changes made to the model geometry. This can be understood with the help of the following formulation. Let the model geometry be represented by $G(p_1 \cdots p_n)$, where $p_1 \cdots p_n$ are constraint parameters; and assume that there is a one-to-one correspondence between parameters and constraints, for simplifying the discussion. Two parameter changes $\delta p_i$ and $\delta p_j$ will yield two model geometry changes $\Delta G_i$ and $\Delta G_j$, where $\Delta G_i = G(p_1 \cdots p_i + \delta p_i \cdots p_n) - G(p_1 \cdots p_i \cdots p_n)$, and the same for $\Delta G_j$. $\Delta G_i/\delta p_i$ and $\Delta G_j/\delta p_j$ are generally not equal, and the constraint corresponding to the larger one has more impact on the model geometry than the other.

For a constraint with a high rate of $\Delta G_i/\delta p_i$, a small parameter change made to it will lead to a large change on the model geometry, which may lead to an unpredictable model variation and a large deviation from the likely original design intent. A model under such a situation is said to have a poor constraining scheme [11]. Because of these, the rate of model geometry change (to be called change rate when the meaning is clear from context) is chosen to quantify a constraint's impact on the model geometry. With this notion, it is expected that resolution options leading to a model with the least change rate will be chosen. Then the least model variations could be expected for later parametric edits.

[2] This table does not include the precedence for compound constraints. A compound constraint's precedence is defined as the highest precedence of its constituent constraints. For example, a tangent constraint between a plan and a cylinder has two constituent constraints: perpendicular directions and distance between positions. Its precedence is then the higher one of these two, which in this case are both 1.

In the discussion above, the model geometry change was loosely denoted by $\Delta G_i$ without a formal definition. We next give a mathematical modeling of this notion. It will begin with comparing constraints in a well-constrained model and then extend the result to general models involving under-constrained, well-constrained, and over-constrained parts. Certainly, there is no need to prioritize constraints in a well-constrained model since no resolution is needed for such a model, but starting with it allows an easier presentation of the modeling.

Recall that a model's GCS can be represented by a system of algebraic equations. This time the system is to be denoted by $F(X, P) = 0$, $P$ being constraint parameters. A nonzero parameter change $\Delta P$ will lead to a change $\Delta X$ made to the model geometry, and they are related by $\frac{\partial F}{\partial X}\Delta X + \frac{\partial F}{\partial P}\Delta P = 0$ where $\frac{\partial F}{\partial X}$ is the Jacobian matrix that was denoted by $J(X)$ in Eq. (1). If we substitute $\frac{\partial F}{\partial X}$ with the geometric perturbation matrix $G$, the term $\Delta X$ will have a clear geometric meaning: geometry change expressed in terms of geometric motions. The relationship between $\Delta P$ and $\Delta X$ then takes the following form:

$$G\Delta X + \frac{\partial F}{\partial P}\Delta P = 0 \Leftrightarrow G\Delta X = -\frac{\partial F}{\partial P}\Delta P \quad (19)$$

Among all geometry changes, we are interested in the one corresponding to a parameter change made merely to one constraint, say the $i$-th constraint. Such a parameter change can be represented by a special $\Delta P$: $\Delta P(i) = 1$ and $\Delta P(j \neq i) = 0$. Solving Eq. (19) with this $\Delta P$ will give the intended geometry change. Also, as $\Delta P(i)$ has the unit magnitude, the solved $\Delta X$ gives directly the change rate for the $i$-th constraint.

It seems that the term $\Delta X$ is a good candidate to define $\Delta G_i$. However, $\Delta X$ is not invariant under rigid-body motions. For this reason, we slightly modify this term, introducing the notion of relative geometry change. $\Delta X$ represents the absolute change made to the model geometry, and its $i$-th component $\Delta X(i)$ describes the absolute change made to the $i$-th geometric entity. We define the relative change $\Delta X'(i)$ for the $i$-th geometric entity as follows:

$$\Delta X'(i) = \frac{1}{\sum_j w_{ij}} \sum_j w_{ij}\big(\Delta X(i) - \Delta X(j)\big) \quad (20)$$

where the weighted sum is taken over all the neighboring geometric entities $j$. If rewritten in a matrix form, Eq. (20) becomes $\Delta X' = R\Delta X$ with the transformation matrix $R$ being:

$$R_{ij} = \begin{cases} 1 & j = i \\ -\frac{w_{ij}}{\sum_j w_{ij}} & j \neq i \ and\ j\ is\ neighboring\ to\ i \\ 0 & otherwise \end{cases} \quad (21)$$

With the notion of relative geometry change in place, $\Delta G_i$ is to be defined as the following quantity:

$$\|\Delta X'\|_2 = \|R\Delta X\|_2 = \left\| RG^{-1}\frac{\partial F}{\partial P}\Delta P_i \right\|_2 \quad (22)$$

The second equality is due to Eq. (19). In the equation, $\Delta P_i$ denotes the vector: $\Delta P(i) = 1$ and $\Delta P(j \neq i) = 0$. One can easily verify that this quantity is invariant under rigid-body motions. In addition, the formulation above has an interesting geometric meaning. In differential geometry, the matrix $R$ is known as the Laplace operator that can be used to measure a model's (discrete) mean curvatures [31,32]. Hence, the quantity in (22) approximately measures how the model's total mean curvature varies with the constraint parameter change. This may partly explain the effectiveness of the proposed constraint prioritization scheme.

Evaluating the change rate for a constraint in an over-constrained model should be done based on the constraints having dependency with it. This is because we want to know the impact of removing this constraint on the model geometry rather than its current impact on the model geometry. In other words, the constraint in a minimal over-constrained part that is preferred to being removed should lead to the smallest summed change rate of the other constraints in the part. Let a minimal over-constrained part be represented by $g$. The evaluation of the change rate for a constraint in $g$, say $c_i$, is to be based on the following quantity:

$$\sum_{c_j \in g, j \neq i} \left\| RG^{-1}\frac{\partial F}{\partial P}\Delta P_j \right\|_2 \quad (23)$$

This time, the term $G^{-1}$ denotes the pseudo inverse of $G$, instead of the ordinary inverse [29]. This is because the matrix $G$ is neither full column rank nor full row rank for general models; the pseudo inverse should be used for such matrices.

For an under-constrained model, the evaluation is performed for constraints to be added to the model rather than the constraints already presented in the model. This is to be done by first virtually adding a resolution option of interest to the model, and then evaluating the change rate for this constraint using the "new" model based on a variant of the quantity in Eq. (22):

$$\left\| RG^{-1}\frac{\partial F}{\partial P}\Delta P_{n+1} \right\|_2 \tag{24}$$

where the term $G^{-1}$ is also the pseudo inverse of $G$, and $n$ is the number of constraints before the virtual addition.

Algorithm 3 summarizes the procedures used to compare two resolution options given by the previous resolution option generation module. Lines 3 and 8 perform the rough comparison, and Lines 5 and 6 the fine comparison. The Boolean variable $FLAG$ is used to control the different priority assignments for over-constraint resolution options and under-constraint resolution options. For over-constraint resolution options, the algorithm outputs the constraint with the lower type precedence or with the same type precedence but a lower summed change rate, while for under-constraint resolution options, it outputs the constraint with the higher type precedence or with the same precedence but a lower change rate.

So far, all the essential modules in the GAI resolution framework presented in Fig. 2 have been made available. Assembling them together as in the framework gives the final overall algorithm of GAI resolution. As the algorithm follows exactly the workflow given in the framework, we do not further present it as an additional algorithm box. With the algorithm, what the user needs to do is just to accept or reject the incrementally presented resolution suggestions; all the decision-support information will be generated automatically by the computer.

**Algorithm 3**: Resolution Option Comparison

**Input:** $c_i, c_j$ − two given resolution options

**Output:** $c$ − the constraint with a higher removal/addition priority

1. **if** $c_i, c_j$ are over-constraint resolution options **then** $FLAG \leftarrow TRUE$ **else** $FLAG \leftarrow FALSE$
2. **if** $c_i$ has a higher type precedence than $c_j$ **then** // use Table 3
3.     **if** $FLAG$ **then** $c \leftarrow c_j$ **else** $c \leftarrow c_i$
4. **else if** $c_i$ has the same type precedence as $c_j$ **then**
5.     **if** $FLAG$ **then** $r_1, r_2 \leftarrow$ Evaluate (23) for $c_i, c_j$ **else** $r_1, r_2 \leftarrow$ Evaluate (24) for $c_i, c_j$
6.     **if** $r_1 > r_2$ **then** $c \leftarrow c_j$ **else** $c \leftarrow c_i$
7. **else**
8.     **if** $FLAG$ **then** $c \leftarrow c_i$ **else** $c \leftarrow c_j$
9. **Return** $c$

## 5. Results

### 5.1. Implementation

The methods presented previously have been implemented using C++ on top of the direct modeling system developed in the authors' previous work [3]. Fig. 3 shows the graphical user interface of the implemented GAI resolution prototype, using QT (version 5.7). All the numeric solving/optimization was carried out using the C++ library Eigen (version 3.2.9) and MATLAB (version R2017a). To start GAI resolution, the user presses the Analyze/Resolve button in the inconsistency resolution toolbox (labeled as 3). The user has the option to let the computer take care of all the work (including inconsistency reasoning and decision-making) by checking the two Auto options in the toolbox. After the activation is done, the computer analyzes the constraint state of the model. If the model if well-constrained, a message box pops up to notify the user of this state; otherwise, a panel pops up to show the inconsistency detection results (labeled as 4) and the generated resolution suggestions (labeled as 5). The suggestions are numbered according to their priorities. The user then chooses among these suggestions.

### 5.2. Case studies and comparisons

The presented methods have been tested using models from both real and simulated data, and some of the results are to be provided. The effectiveness of the detection module will be shown by comparisons with existing methods.

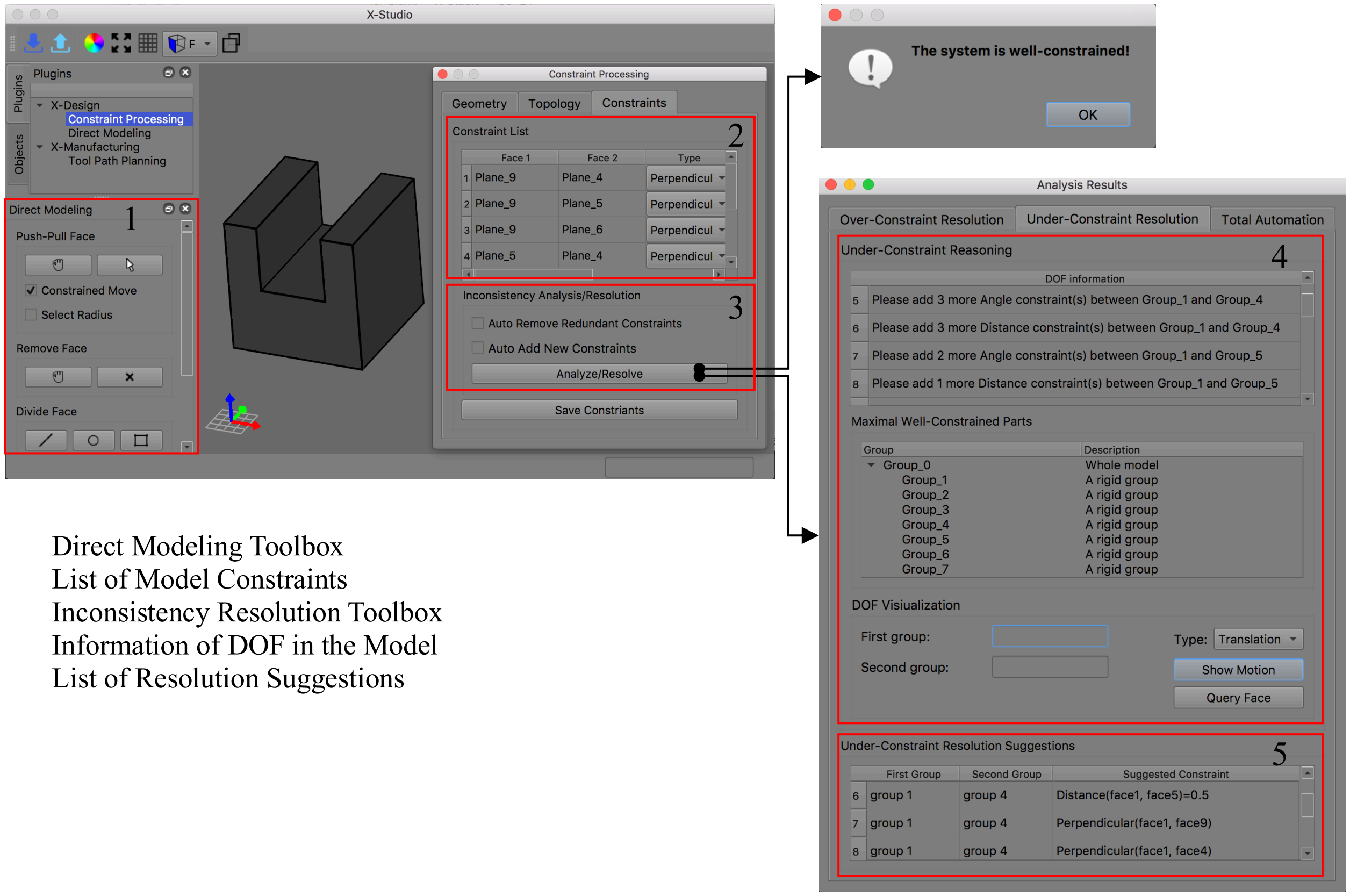

Figure 3: Graphical user interface of the GAI resolution prototype.

The effectiveness of the prioritization module, in particular the change rate notion, will be demonstrated first through two simple examples and then via two comprehensive cases that are based on real-world mechanical parts.

Fig. 4 shows the under-constrained model[3] used to demonstrate comparisons between the proposed method and the previous greedy method [17] for maximal well-constraint detection. The comparison results are depicted in Fig. 5. The greedy method did not give the optimal result, while the proposed method was able to. This confirms our previous statement that the greedy method could generate undesirable detection results and only represent an incomplete technical tool rather than an appropriate formulation of the detection problem. A similar situation occurred for minimal over-constraint detection (Fig. 6). The over-constraint is a consequence of the dependency among constraints C1, C2, C5 and C7; and constraint C8 is a duplication of constraint C7. Even for this simple case, the greedy method gave wrong detection results (Fig. 6c). The greedy method uses the following simplified procedures to conduct minimal over-constraint detection (Section 3, [17]): (1) begin with a seed constraint, say C1; (2) then iterate through all constraints to greedily find a maximal subset of independent constraints, which is {C1-C6} for this case; (3) finally check the dependencies of the constraints not presented in the subset with the constraints in the subset, resulting in the two over-constrained parts shown in Fig. 6c. However, the model has an over-constrained part with a smaller size as shown by Part II in Fig. 6b. The reason for the greedy method's ineffectiveness is that the maximal independent subset is not unique, and this non-uniqueness could lead to wrong detection results.

In the prioritization module, the core part is the change rate notion. Figs. 7 and 8 show the application results of this notion, with the weights in Eq. 21 being $w_{ij} = 1$. According to the results, we prioritized the resolution options, and if the two top options are chosen, the two models become well-constraint immediately. The resulting constraining schemes are also those commonly used in practice. These two examples are fairly simple. So another two complex examples were conducted to further show the effectiveness of this notion, as well as the whole decision-support method. One of them is based on a crank model (Fig. 9a), and the other is based an engine bracket model (Fig. 12a).

---

[3] In this model and the later crank model, the plane-cylinder distance constraint refers to the distance from the cylinder's position to the plane.

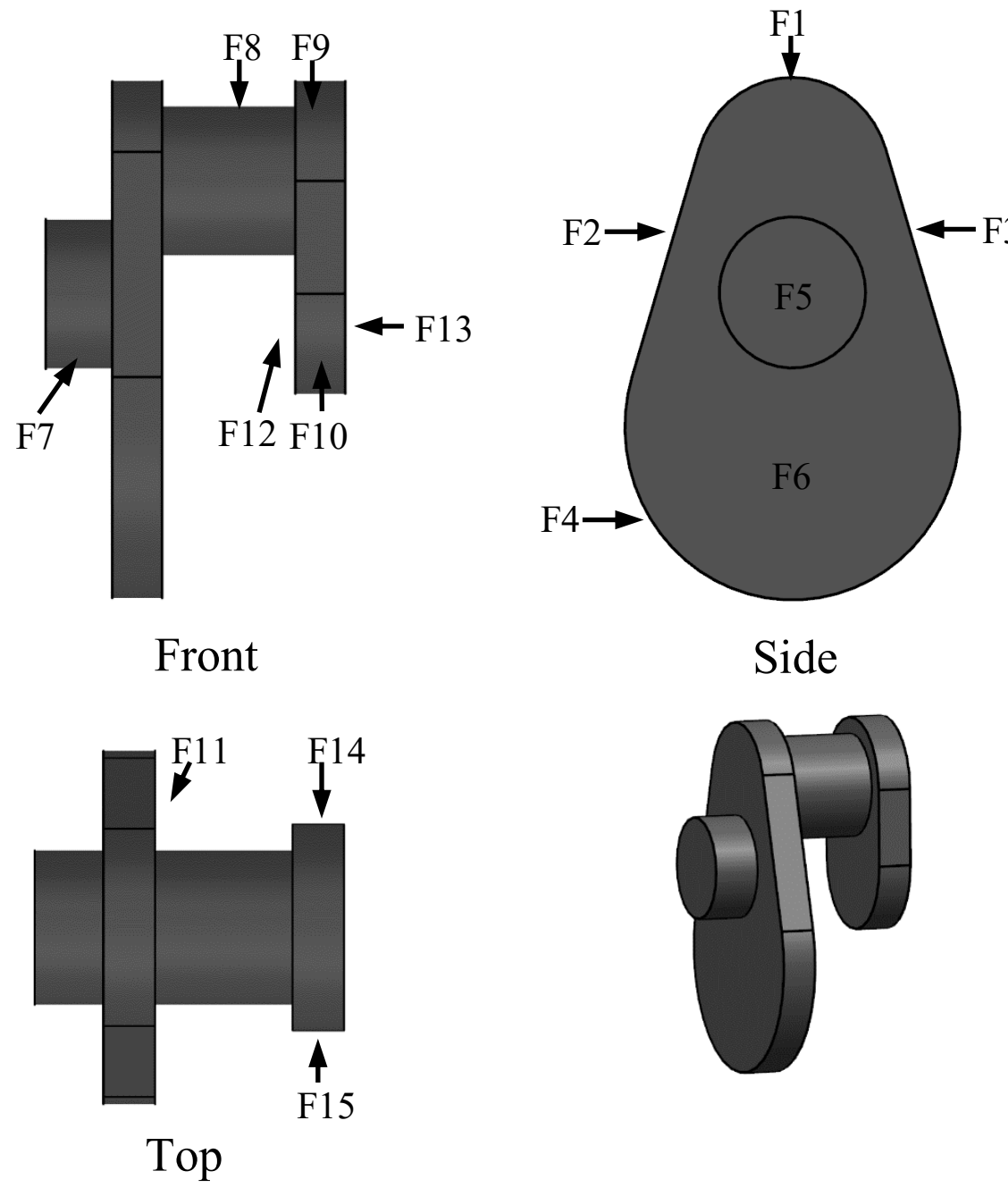

Distance(F1,F7)=34
Distance(F4,F7)=40
CoplanarAxis(F1,F4,F7)
Coaxial(F1,F8,F9)
Coaxial(F7,F10)
Tangent(F4,F2)
Tangent(F4,F3)
Tangent(F10,F14)
Distance(F9,F14)=30
Tangent(F10,F15)
Distance(F9,F15)=30
Perpendicular(F6,F7)
Distance(F5,F6)=20
Distance(F6,F11)=15
Distance(F11,F12)=40
Distance(F12,F13)=15

Figure 4: An under-constrained model.

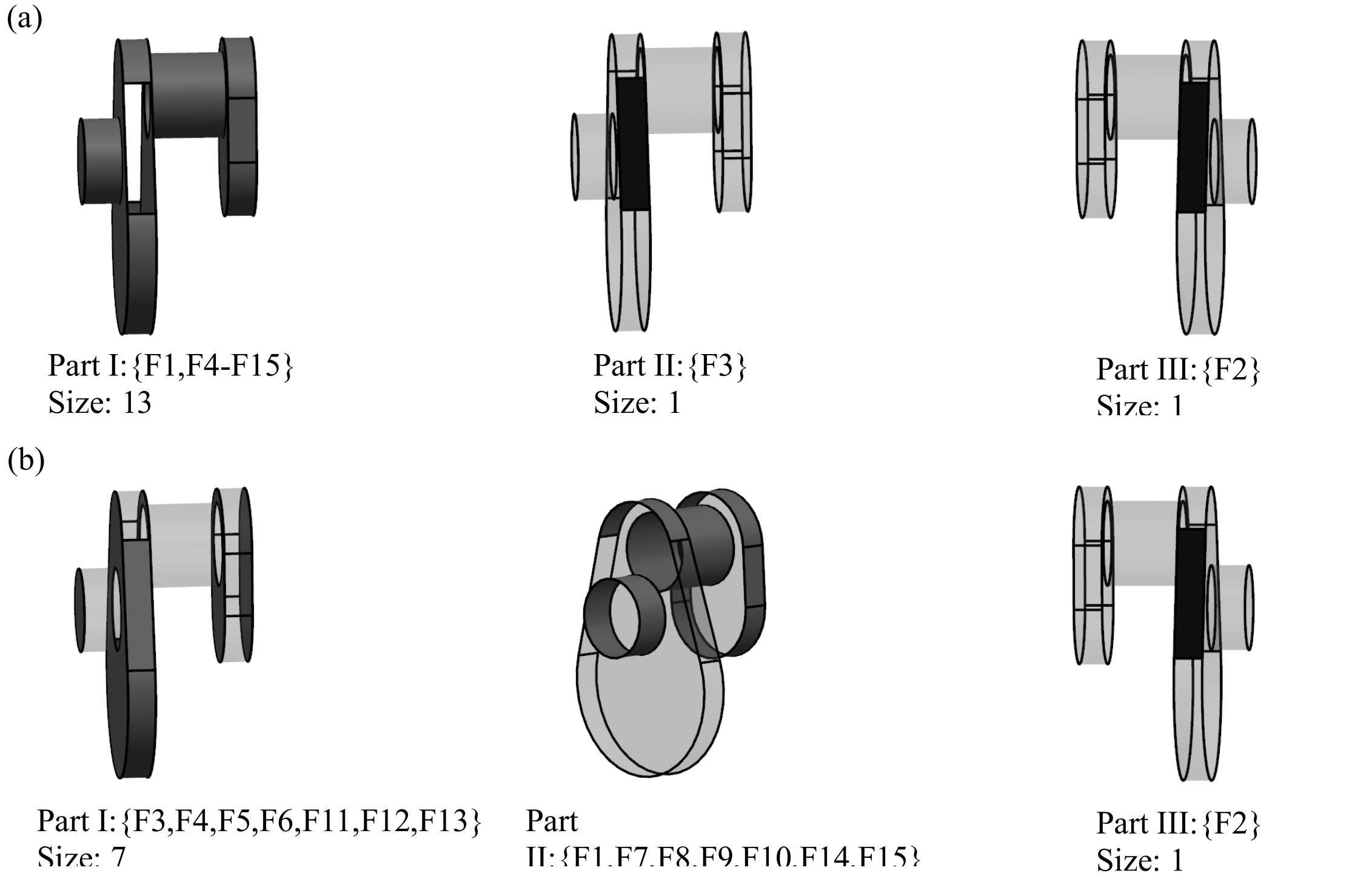

Figure 5: Comparison results of maximal well-constraint detection: (a) the proposed method; and (b) the greedy method.

The direct edit involved in the crank-based case study is depicted in Fig. 9b, which translated the blue faces along the directions indicated by the red arrows. The changes made to the model geometry were the removal of faces F2 and F3, which in turn led to the removal of constraints C6, C7 and C10-C13 in the model GCS, as shown by the GCS update in Fig. 9c. To simplify the presentation, constraints for the model's right part (in the dashed rectangle, Fig.9a) were omitted, as this part is the same as the left part. The GCS update resulted in an under-constrained model, a result given by the analyzer module in the framework. The detection module found that there were three maximal well-constrained parts in the model, which are labeled as Maximal Well-Constrained Parts I, II and III in Fig. 10. In this figure, the resolution options generated by the prioritization module are also shown, in the form of two lists of suggestions. Due to the large number of resolution suggestions produced, we do not present them in detail; only the primary suggestions and those to be referenced in later discussions are presented.

The DOF between part I and part II is a rotation of face F5 about the axis of face F6. Each of the constraints in the left suggestion list can eliminate this DOF, and the user has the discretion to add which of them. In particular, if the user lets the computer take care of everything, the top suggested constraint will be used (in the red rectangle). Once a suggested constraint is selected, there is no DOF between part I and part II anymore; then, these two parts will be merged to form a new maximal well-constrained part, as shown by the model labeled as Merged Maximal Well-Constrained Part. Following the same procedures as just described, the DOFs between this merged part and part III can be eliminated, and then a well-constrained crank model is attained.

(a)

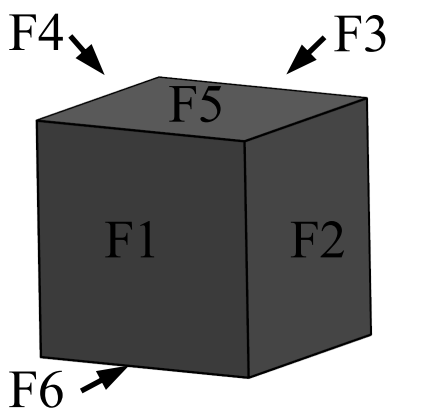

Distance(F1,F3)=1
Distance(F2,F4)=1
Distance(F5,F6)=1
Perpendicular(F1,F5)
Perpendicular(F1,F4)
Perpendicular(F4,F5)
Perpendicular(F2,F3)
Perpendicular(F2,F3)

(b)

Part I: {C1,C2,C5,C7}
Part II: {C7,C8}

(c)

Part I: {C1,C2,C5,C7}
Part II: {C1,C2,C5,C8}

Figure 6: Comparison results of minimal over-constraint detection: (a) an over-constrained model; (b) the proposed method; and (c) the greedy method.

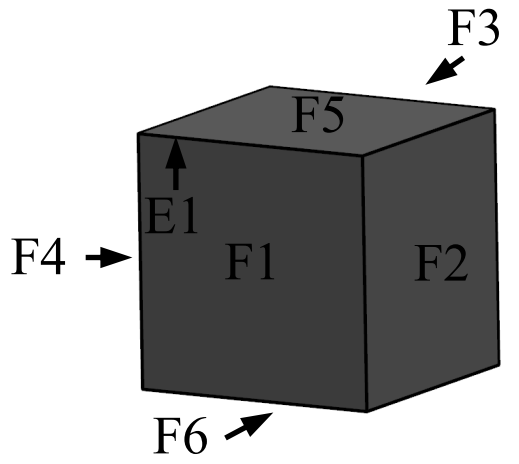

Distance(F1,F3)=1
Distance(F5,F6)=1
Perpendicular(F1,F5)
Perpendicular(F1,F4)
Perpendicular(F4,F5)
Parallel(F2,F4)
Perpendicular(F2,F5)
Length(E1)=1

| Resolution Option | Summed Rate |
|---|---|
| Remove C5 | 2.2500 |
| Remove C7 | 2.2500 |
| Remove C6 | 3.1250 |

Figure 7: Application results of the change rate notion to an over-constraint model.

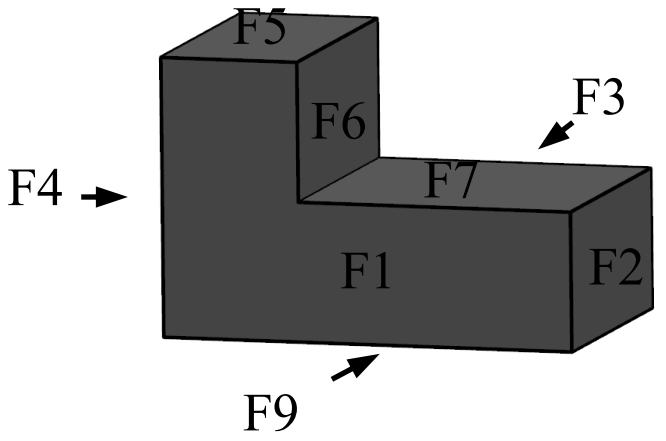

Distance(F1,F3)=10
Distance(F2,F4)=15
Distance(F5,F9)=10
Perpendicular(F1,F2)
Perpendicular(F1,F9)
Perpendicular(F2,F9)
Distance(F5,F7)=5

| Resolution Option | Change Rate |
|---|---|
| Add Distance(F4,F6) | 0.0258 |
| Add Distance(F2,F6) | 0.0258 |
| Add Perpendicular(F6,F3) | 1.0172 |
| Add Perpendicular(F6,F1) | 1.0172 |
| Add Perpendicular(F6,F5) | 1.1806 |
| Add Perpendicular(F6,F7) | 1.1806 |
| Add Perpendicular(F6,F9) | 1.1806 |
| ⋮ | ⋮ |

Figure 8: Application results of the change rate notion to an under-constraint model.

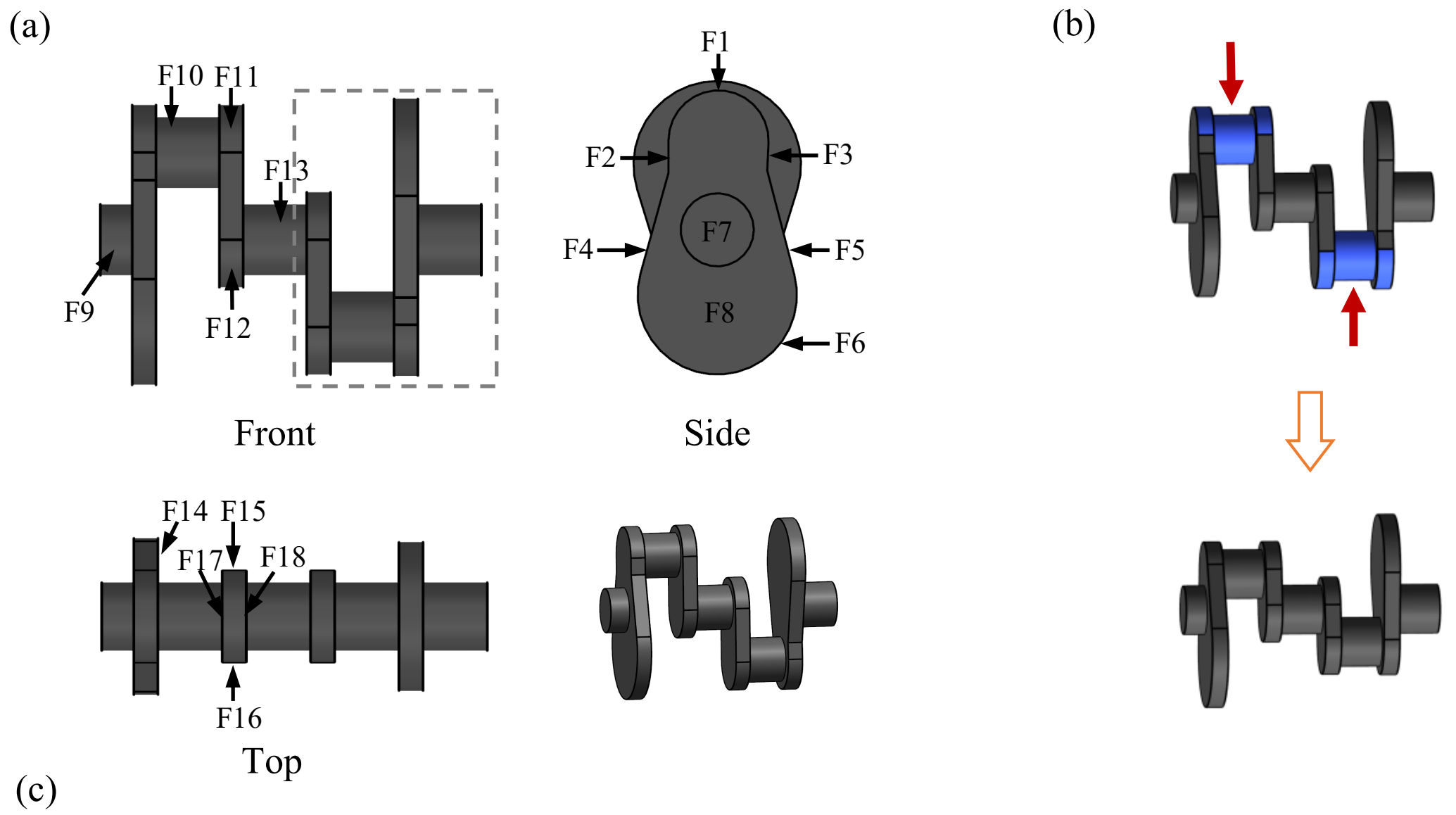

(c)

| Original Constraints | | | | Updated Constraints | | | |
|---|---|---|---|---|---|---|---|
| C1. | Distance(F1,F9)=55 | C12. | Distance(F2,F9)=30 | C1. | Distance(F1,F9)=34 | C12. | Removed |
| C2. | Distance(F6,F9)=40 | C13. | Distance(F3,F9)=30 | C2. | Distance(F6,F9)=40 | C13. | Removed |
| C3. | CoplanarAxis(F1,F6,F9) | C14. | Tangent(F15,F11) | C3. | CoplanarAxis(F1,F6,F9) | C14. | Tangent(F15,F11) |
| C4. | Coaxial(F1,F10,F11) | C15. | Distance(F15,F12)=30 | C4. | Coaxial(F1,F10,F11) | C15. | Distance(F15,F12)=30 |
| C5. | Coaxial(F9,F12,F13) | C16. | Tangent(F16,F11) | C5. | Coaxial(F9,F12,F13) | C16. | Tangent(F16,F11) |
| C6. | Tangent(F1,F2) | C17. | Distance(F16,F12)=30 | C6. | Removed | C17. | Distance(F16,F12)=30 |
| C7. | Tangent(F1,F3) | C18. | Perpendicular(F8,F6) | C7. | Removed | C18. | Perpendicular(F8,F6) |
| C8. | Tangent(F6,F5) | C19. | Distance(F7,F8)=20 | C8. | Tangent(F6,F5) | C19. | Distance(F7,F8)=20 |
| C9. | Tangent(F6,F4) | C20. | Distance(F8,F14)=15 | C9. | Tangent(F6,F4) | C20. | Distance(F8,F14)=15 |
| C10. | Angle(F2,F4)=16.92° | C21. | Distance(F14,F17)=40 | C10. | Removed | C21. | Distance(F14,F17)=40 |
| C11. | Angle(F3,F5)=16.92° | C22. | Distance(F17,F18)=15 | C11. | Removed | C22. | Distance(F17,F18)=15 |

Figure 9: A crank model: (a) face indexing; (b) a direct edit; and (c) GCS update.

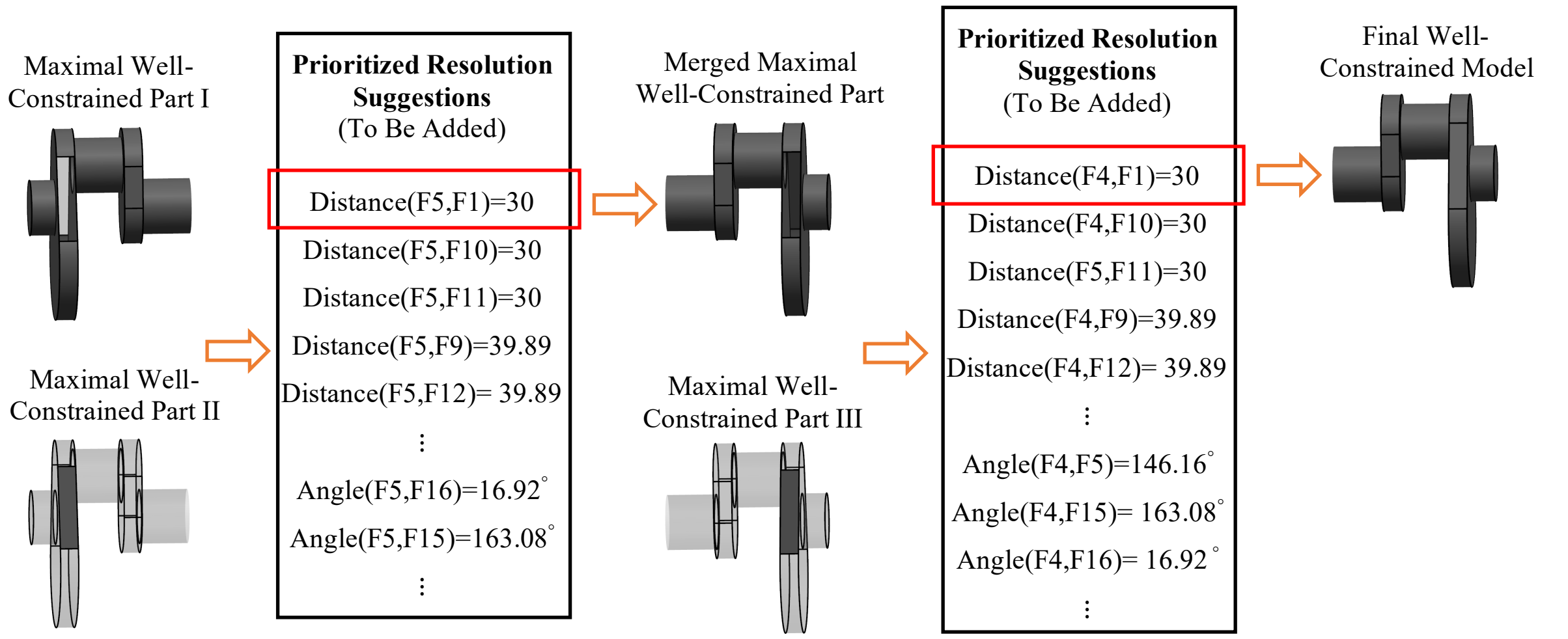

Figure 10: Resolution flow, maximal well-constrained parts, and resolution options/suggestions for the crank model.

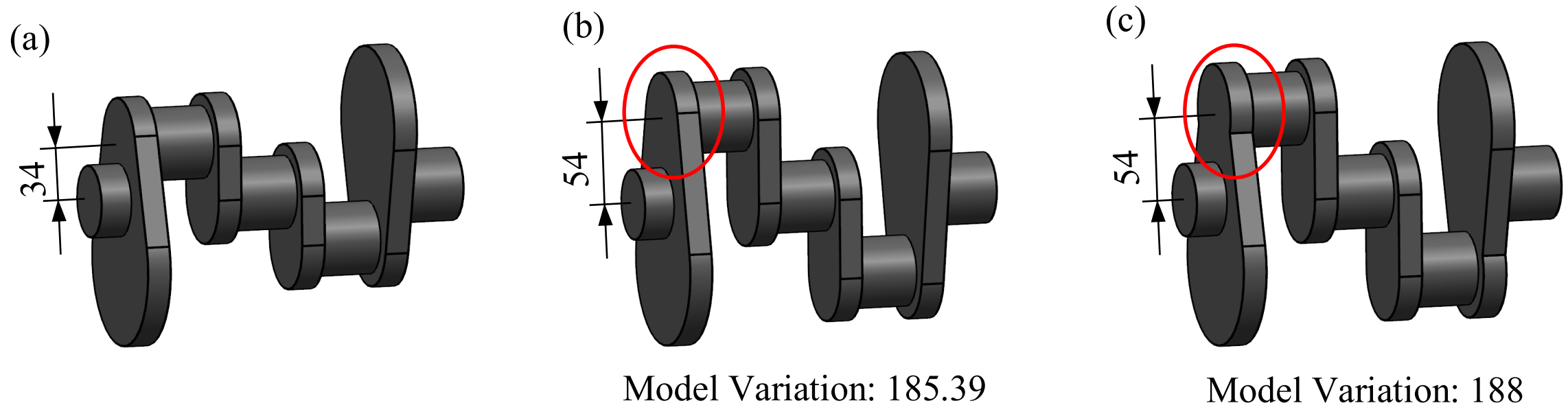

Figure 11: Increase the dimension leading to varied modeling behavior under different resolution decisions for the crank model.

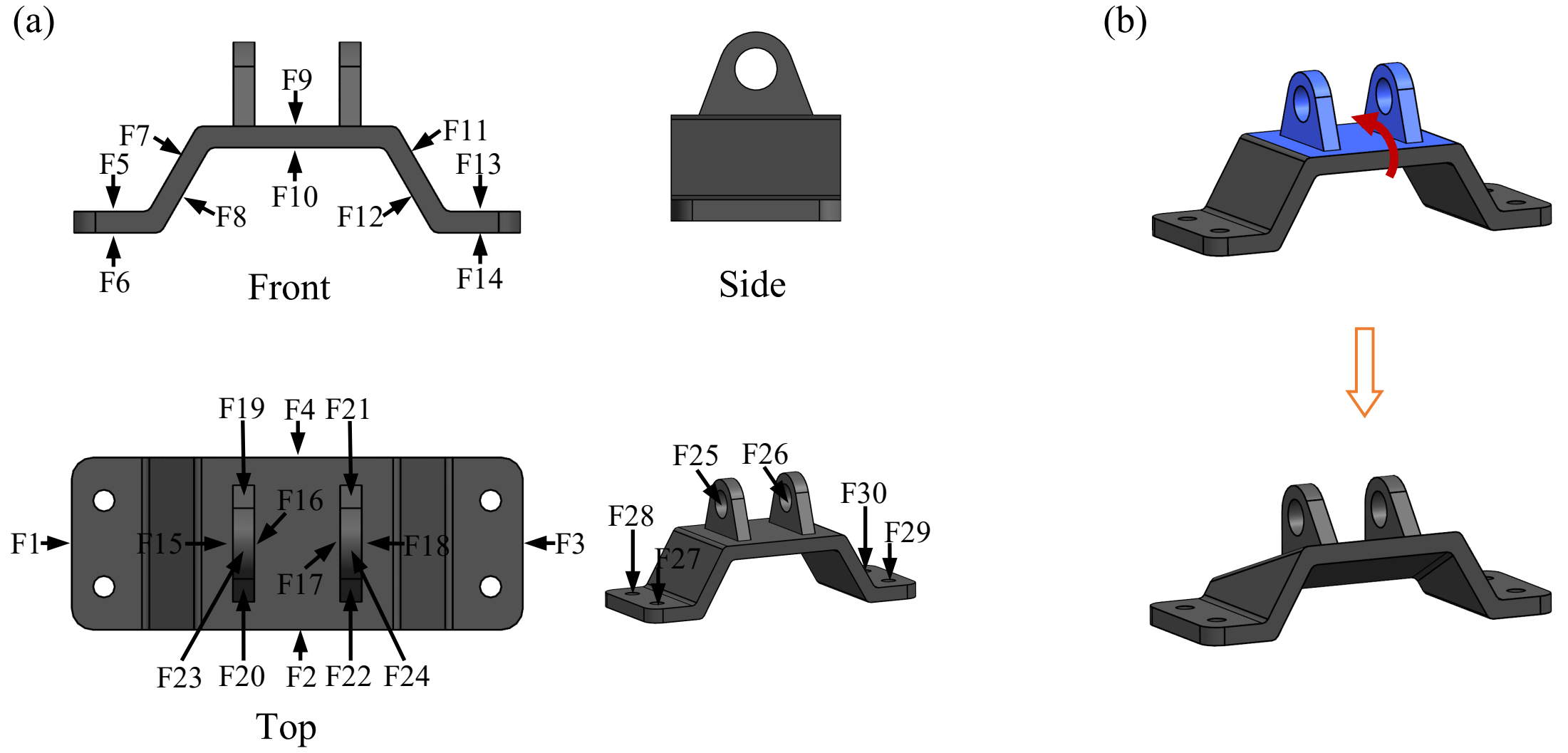

(c)

| Original Constraints | | Updated Constraints | |
|---|---|---|---|
| C1. Dis(F1,F3)=210 | C24. Dis(F15,F16)=10 | C1. Dis(F1,F3)=210 | C24. Dis(F15,F16)=10 |
| C2. Dis(F2,F4)=80 | C25. Dis(F17,F18)=10 | C2. Dis(F2,F4)=80 | C25. Dis(F17,F18)=10 |
| C3. Dis(F5,F6)=10 | C26. Coaxial(F23,F25) | C3. Dis(F5,F6)=10 | C26. Coaxial(F23,F25) |
| C4. Dis(F7,F8)=10 | C27. Dis(F14,F24)=72 | C4. Dis(F7,F8)=10 | C27. Dis(F14,F24)=56.99 |
| C5. Dis(F9,F10)=10 | C28. Dis(F2,F24)=40 | C5. Dis(F9,F10)=10 | C28. Dis(F2,F24)=45.10 |
| C6. Dis(F11,F12)=10 | C29. Coaxial(F24,F26) | C6. Dis(F11,F12)=10 | C29. Coaxial(F24,F26) |
| C7. Dis(F13,F14)=10 | C30. Tangent(F23,F19) | C7. Dis(F13,F14)=10 | C30. Tangent(F23,F19) |
| C8. Per(F1,F2) | C31. Tangent(F23,F20) | C8. Per(F1,F2) | C31. Tangent(F23,F20) |
| C9. Per(F2,F6) | C32. Tangent(F24,F21) | C9. Per(F2,F6) | C32. Tangent(F24,F21) |
| C10. Per(F1,F6) | C33. Tangent(F24,F22) | C10. Per(F1,F6) | C33. Tangent(F24,F22) |
| C11. Per(F2,F8) | C34. Ang(F19,F4)=20° | C11. Per(F2,F8) | C34. Par(F19,F4) |
| C12. Per(F2,F10) | C35. Ang(F20,F2)=20° | C12. Ang(F2,F10)=110° | C35. Ang(F20,F2)=40° |
| C13. Per(F2,F12) | C36. Ang(F21,F4)=20° | C13. Per(F2,F12) | C36. Par(F21,F4) |
| C14. Coplanar(F6,F14) | C37. Ang(F22,F2)=20° | C14. Coplanar(F6,F14) | C37. Ang(F22,F2)=40° |
| C15. Ang(F1,F7)=30˚ | C38. Dis(F27,F1)=10 | C15. Ang(F1,F7)=30° | C38. Dis(F27,F1)=10 |
| C16. Ang(F3,F11)=30˚ | C39. Dis(F27,F2)=20 | C16. Ang(F3,F11)=30° | C39. Dis(F27,F2)=20 |
| C17. Dis(F1,Edge(F6,F8))=40.77 | C40. Dis(F28,F1)=10 | C17. Dis(F1,Edge(F6,F8))=40.77 | C40. Dis(F28,F1)=10 |
| C18. Dis(F3,Edge(F12,F14))=40.77 | C41. Dis(F28,F4)=20 | C18. Dis(F3,Edge(F12,F14))=40.77 | C41. Dis(F28,F4)=20 |
| C19. Dis(F6,F23)=72 | C42. Dis(F29,F3)=10 | C19. Dis(F6,F23)= 56.99 | C42. Dis(F29,F3)=10 |
| C20. Dis(F2,F23)=40 | C43. Dis(F29,F2)=20 | C20. Dis(F2,F23)=45.10 | C43. Dis(F29,F2)=20 |
| C21. Dis(F10,F23)=32 | C44. Dis(F30,F3)=10 | C21. Dis(F10,F23)=32 | C44. Dis(F30,F3)=10 |
| C22. Dis(F1,F15)=75 | C45. Dis(F30,F2)=20 | C22. Dis(F1,F15)=75 | C45. Dis(F30,F2)=20 |
| C23. Dis(F3,F18)=75 | | C23. Dis(F3,F18)=75 | |

Figure 12: An engine bracket model: (a) face indexing; (b) a direct edit; and (c) GCS update (Dis: Distance, Ang: Angle, Per: Perpendicular, Par: Parallel, Tan: Tangent).

**Minimal Over-Constrained Part I**

C2: Distance(F2,F4)=80
C20: Distance(F2,F23)=45.10
C30: Tangent(F23,F19)
C34: Parallel(F19,F4)

⇨

**Prioritized Resolution Suggestions**
(To Be Removed)

C34: Parallel(F19,F4)
C30: Tangent(F23,F19)
C20: Distance(F2,F23)=45.10
C2: Distance(F2,F4)=80

**Minimal Over-Constrained Part II**

C2: Distance(F2,F4)=80
C28: Distance(F2,F24)=45.10
C32: Tangent(F24,F21)
C36: Parallel(F21,F4)

⇨

**Prioritized Resolution Suggestions**
(To Be Removed)

C36: Parallel(F21,F4)
C32: Tangent(F24,F21)
C28: Distance(F2,F24)=45.10
C2: Distance(F2,F4)=80

Figure 13: Minimal over-constrained parts and resolution options/suggestions for the braket model.

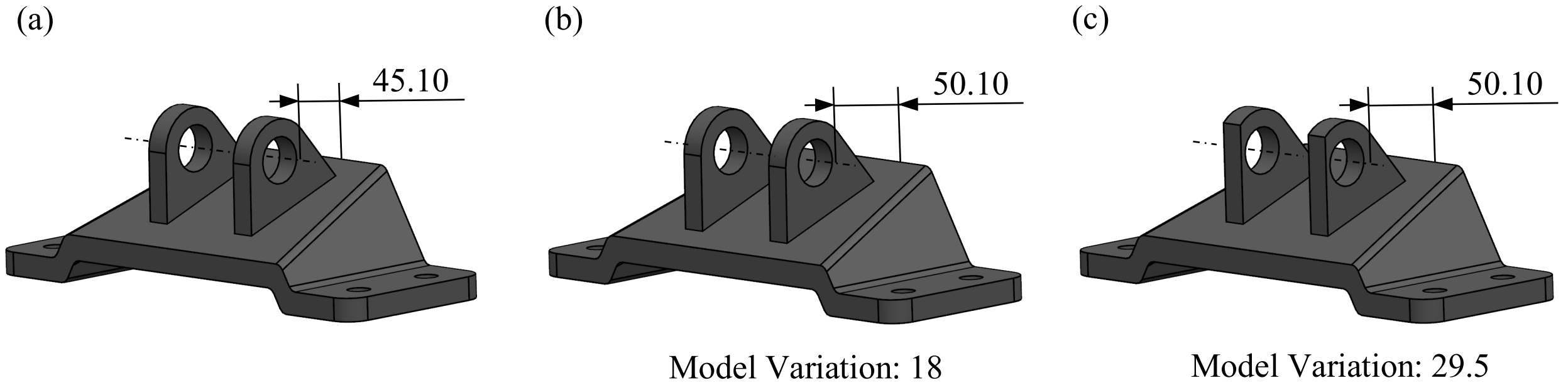

Figure 14: Increase the dimension leading to varied modeling behavior under different resolution decisions for the bracket model.

Different selections of suggested resolution options will lead to different modeling behavior under parametric edits. Fig. 11 shows the varied modeling behavior of the crank model for different resolution decisions. The tested parameter change was made to a key dimension that relates all of the three maximal well-constrained parts. Fig. 11b shows the result of choosing the top suggested constraints; Fig. 11c shows the result of choosing two other suggested constraints: $Angle(\text{F5}, \text{F16}) = 16.92°$ and $Angle(\text{F4}, \text{F5}) = 146.16°$. As expected, the resolution decision whose constraints occurred first in the recommendation lists yielded a less model variation value. From an intuitive point of view, the model's shape in Fig. 11a is preserved better in the model of Fig. 11b than in that of Fig. 11c, which can demonstrate the effectiveness of the change rate notion.

The case study above involves merely under-constraint, and thus it allows us to focus on the validation of modules in the under-constraint resolution branch in the framework. This, on the other hand, implies that modules in the other branch were not executed. For this reason, the bracket-based case study was conducted. The applied direct edit and the associated GCS update for this case study are shown in Fig. 12. These changes led to two minimal over-constrained parts {C2,C20,C30,C34} and {C2,C28,C32,C36}. Due to the symmetry of the model, these two parts have similar configurations. We thus only need to focus on one of them, and the other can be resolved with the same process. Take part I as an example. The prioritization of its four resolution options is shown in the mid-left rectangle in Fig. 13. As in the previous case study, the user has the discretion to remove any of them and, if the user decides not to control the resolution, the top suggested constraint will be removed. Once a suggested constraint is selected and removed, the cyclic dependency in the group under resolution is broken, and the associated over-constraint is resolved.

Fig. 14 shows the modeling behavior of different resolution decisions under a same parametric edit. Fig. 14b shows the result of choosing the top suggested constraints, i.e., removing constraints C34 and C36. Fig. 14c shows the result of choosing two other suggested constraints: removing constraints C30 and C32. As can be seen from the model variation values in Fig. 14b and Fig. 14c, the proposed methods, again, assigned a higher priority to the resolution suggestions that can lead to a less model variation and a better resemblance to the original model. It should be noted here that the two decisions behind Fig. 14b and Fig. 14c resulted in under-constraint in the model, which were resolved by adding constraints $Angle(\text{F19}, \text{F9}) = 70°$, $Angle(\text{F21}, \text{F9}) = 70°$, and $Distance(\text{F19}, \text{F4}) = 18.71$, $Distance(\text{F21}, \text{F4}) = 18.71$, respectively.

### 5.3. Limitations and discussion

The varied modeling behavior shown in Fig. 11 and Fig. 14 has been presented to three experienced mechanical engineers in the university's machine shop. They all chose Fig. 11b and Fig. 14b over those depicted in Fig. 11c and Fig. 14c, which can partly confirm the effectiveness of the presented decision-support method. Nevertheless, two of them further pointed out that Fig. 11c was seen in practice, although quite uncommon. This basically implies that it may not be possible to have a general prioritization criterion that can always give the most intended resolution option, at least for general CAD systems. Hence, for GAI resolution, the automatic mode should be working as a secondary mode, and the decision-support mode should be a primary way of working in practice.

Although the presented decision-support method is seen to be quite effective in the case studies above, it is expected that the ambiguity among resolution options will increase in more complicated cases. As a result, the prioritization results may deviate from the user's design intent more, and the user may need to go through more resolution suggestions before getting the right constraint(s). There are hints for such a situation in the crank-based case study. The first three options in the left suggestion list of Fig. 10 have the same change rate value (= 0.9517), meaning that the change rate notion cannot distinguish them. But they clearly represent different design intents. This implies that the change rate notion is not in line with the design intent perfectly, partly because this notion only characterizes the local model variation behavior under parametric edits without considering the global behavior.

This work's research so far suggests that the presented decision-support method can be used to generate resolution options with guaranteed validity and to guide the user's choice among these resolution options. The change rate notion is a good quantification of a constraint's impact on the model geometry, but it does not match the design intent perfectly. This gap may be improved by priorities incorporating both local and global characterizations of model variation. Modeling a model's global variation behavior is, however, very challenging due to the nonlinearity issues involved.

## 6. Conclusions

Direct modeling is a very recent CAD paradigm that has been recognized by industry. It, however, lacks the parametric capability. The major issue to enable this capability has been found to be the possible information inconsistencies in a model after direct edits and the many inconsistency resolution options. The challenge lies in effectiveness towards detecting minimal over-constrained parts and maximal well-constrained parts in the model and an effective criterion for prioritizing valid resolution options. New/improved methods have thus been presented in this work to solve this challenge. In particular, a new, precise formulation of the detection problem was proposed; a hybrid prioritization criterion combining the existing type-based criterion and a newly proposed change rate based criterion was presented. They together led to an implementation of the decision-support framework outlined in Fig. 2. The practical benefits of this framework include an automatic inconsistency reasoning, a complete avoidance of invalid resolution options, and a guided exploration of valid resolution options. The developed methods have been validated with a series of case studies and comparisons.

A couple of limitations need to be noted here. As the present work only formulated a model's local variation behavior, decision-making that requires additional insights into the model's global variation behavior may not be supported satisfactorily. A more comprehensive prioritization criterion considering both local and global variation behavior could address this issue. Nevertheless, modeling the global variation behavior introduces the very challenging nonlinear issues. Machine learning algorithms may serve to address the issues because they can treat nonlinearity as the black box. It should also be noted that resolution options are currently presented to the user via plain text. Animations simulating the model variation behavior under different resolution options can be very practically beneficial, and are among the CAD research studies to be carried out in our research group.

## Acknowledgements

This work has been funded in part by the Natural Sciences and Engineering Research Council of Canada (NSERC) under the Discovery Grants program.